\documentclass[reprint,amsmath,amssymb,aps,prd,nofootinbib]{revtex4-2}

\usepackage{graphicx}
\usepackage{dcolumn}
\usepackage{bm}
\usepackage[dvipsnames]{xcolor}
\usepackage{ulem}
\usepackage{natbib}

\newcommand{\mpl}{E_{\rm pl}}

\newcommand{\aap}{Astron.\ Astrophys.}
\newcommand{\mnras}{Mon.\ Not.\ R.\ Astron.\ Soc.}

\newcommand{\apjl}{Astrophys.\ J.\ Lett.}

\newcommand{\jcap}{J. Cosmology \& Astropaticles.}

\newcommand{\eqg}{{\cal E}_{\rm QG}}
\newcommand{\eqgone}{{\cal E}_{\rm QG,1}}
\newcommand{\eqgtwo}{{\cal E}_{\rm QG,2}}
\newcommand{\eqgn}{{\cal E}_{\rm QG,n}}

\begin{document}


\title{Lorentz Invariance Violation Limits from GRB 221009A}%

\author{Tsvi Piran}
\email{Tsvi.Piran@mail.huji.ac.il}
\author{Dmitry D. Ofengeim}%
\affiliation{Racah Institute of Physics The Hebrew University, Jerusalem 91904, Israel}

\date{\today}

\begin{abstract}
It has been long conjectured that  a  signature of Quantum Gravity will be Lorentz Invariance Violation (LIV) that could be observed at energies much lower than the Planck scale.
One possible signature of LIV  is an energy-dependent speed of photons. This can be tested with a distant transient source of very high-energy photons.  We explore time-of-flight limits on LIV derived from LHAASO's observations of tens of thousands of TeV photons from GRB 221009A, the brightest gamma-ray burst of all time. For a linear ($n=1$) dependence of the photon velocity on  energy, we find a lower limit on the subluminal (superluminal) LIV scale of  ${5.9} (6.2) \ \mpl$. These are comparable to the stringent limits obtained so far and as an independent bound obtained from a different redshift confirm their robustness. For a quadratic model ($n=2$, corresponding to $d=6$ SME operators), the limits, which are currently  
the  best available 
with the time-of-flight method,
are  $5.8 (4.6) \times 10^{-8}\ \mpl$. 
Our analysis uses the publicly available LHAASO data which is only in the $0.2-7$ TeV range. Higher energy data would enable us to improve these limits by a factor of 3 for $n=1$ and by an order of magnitude for $n=2$.
\end{abstract}

\maketitle

\section{Introduction}
\label{sec:Introduction}
The Gamma-Ray Burst 
(GRB)
221009A was an extremely powerful nearby ($z=0.151$) burst,
the brightest ever detected.  Among its numerous unique features was the detection of tens of thousands of TeV ($0.2-13\,$TeV) photons~\cite{LHAASO2023,LHAASOSci2023_13TeV}. The large number of photons enabled the LHAASO collaboration to obtain detailed TeV lightcurve and spectra. The TeV lightcurve, which  was very different from the lower energy gamma-rays lightcurve detected by Fermi \cite{Fermi2023}, HXMT \cite{HXMT} and other detectors, confirmed earlier suggestions that the dominant GRBs' very high energy emission is an afterglow \cite{Fan2008, Kumar2009,Kumar2010,Ghisellini2010,Derishev2019,Magic2019} and that the prompt very high energy emission is much weaker.  

This TeV lightcurve and spectra enable us to explore the possibility  of  Lorentz Invariance Violation (LIV) and set limits on the scale in which this may happen. LIV was introduced as a potential low-energy signature of Quantum Gravity-induced corrections to general relativity. 
Historically, the concept was suggested as a potential explanation for the GZK paradox \cite{Coleman1999,Gonzalez-Mestres1997} by introducing   a shift in the 
Delta resonance $p+\gamma \rightarrow \Delta^+$  threshold, preventing energy losses of the protons on interaction with the CMB background photons. Later on, it was pointed out that this modification  would also introduce a shift in the annihilation threshold of high energy  photons by the extragalactic background light (EBL),
$\gamma\gamma \rightarrow e^+ e^-$ \cite{Kifune1999,Kluzniak}.  
When an 18 TeV photon was reported to have arrived from GRB 221009 \cite{Huang2022}, LIV has been suggested, among several other `new physics' solutions  \cite{MA}. However, careful processing \cite{LHAASOSci2023_13TeV} of the LHAASO data resulted in lowering the highest detected photon energy to $10-13\,$TeV, relaxing the problem of the  optical depth for this GRB.

LIV also introduces an energy-dependent photon  velocity. This can be tested by comparing the arrival times of different energy photons that have been emitted simultaneously~\cite{Amelino1998}. For this, one needs a distant high-energy transient source with rapid temporal variability. GRBs are the best transients for that. A high fluence distant GRB with a  very high energy emission is the best source to explore LIV-induced time-of-flight (TOF) differences: GRB221009A is an ideal source. 

We  explore here the implication of the rapid rise time of the TeV afterglow to TOF LIV limits. We begin in  \ref{sec:LIV} with a brief discussion of Quantum Gravity-modified dispersion relations. We describe in \ref{sec:analysis} the TeV afterglow data and our LIV limits. We compare in \ref{sec:090510}  our results to earlier LIV limits obtained with  GRBs 090510A and 190114C. We conclude, summarizing our findings, in \ref{sec:conclusions}. 

\section{Lorentz Invariance Violation}
\label{sec:LIV}
A natural expectation for a ``low-energy''  quantum gravity effect is a modified photon dispersion relation  \cite{Amelino1998}:
\begin{equation}
{E^2 = p^2 c^2\left[ 1 
{+\sigma} 
\left(\frac{E}{{\eqgn^{(\sigma)}} \mpl}\right)^n \right]} \ , 
\label{eq:dispersion}
\end{equation}
where  $\mpl \approx 1.22 \times 10^{19}\,$GeV is the Planck energy. $\eqgn^{(\sigma)}$  measures the LIV energy scale in units of $\mpl$ and the index $\sigma = \pm 1$ denotes the sign of the dispersion modification. Such a dispersion introduces an energy-dependent photon velocity. For $E\ll \eqgn \mpl$ we have:
\begin{equation}
\label{eq:v}
v= c \left[1 +\sigma\frac{n+1}{2} \left(\frac{E}{\eqgn^{(\sigma)} \mpl}\right)^n \right]. 
\end{equation}
Subluminal (superluminal) motion corresponds to $\sigma = -1 (+1)$. This leads to an energy-dependent TOF of a photon with energy $E$ as compared to travel at the speed of light \cite{Jacob2008}:
\begin{multline}
\delta t_{\rm LIV} (E)= {-\sigma} \frac{n+1}{2 H_0} \left(\frac{E}{\eqgn^{{(\sigma)}} \mpl}\right)^n \\
\times \int_0^z \frac{(1+\zeta)^n d\zeta}{\sqrt{\Omega_{\rm m} (1+\zeta)^3+\Omega_\Lambda}} \ , 
\label{eq:TOF}
\end{multline}
where $z$ is the redshift of the source, $H_0= 67.5\,$km$\,$s$^{-1}\,$Mpc$^{-1}$ is the Hubble-Lemaitre constant, $\Omega_{\rm M}=0.315$ is the current fraction of matter in the Universe, and $\Omega_\Lambda=0.685$ the  cosmological constant fraction. For GRB 221009A at $z=0.151$ we have:
\begin{equation} 
\delta t_{{\rm LIV},z=0.151}(E) = {\sigma} 
\begin{cases}
			5.7\,\text{s}\, \dfrac{E/{\rm TeV}}{\eqgone^{{(\sigma)}} }, & {n=1}\\
        	7.5\,\text{s}\,  \left( \dfrac{E/{\rm TeV}}{10^8\eqgtwo^{{(\sigma)}}} \right)^2, & {n=2}\ .
\end{cases}
\label{eq:TF}
\end{equation}

\section{LIV limits from GRB 221009A}
\label{sec:analysis}
\subsection{GRB 221009A---the brightest burst ever}
The extreme brightness of GRB~221009A was due to a combination of an extreme intrinsic luminosity, $E_{\rm iso} \approx  10^{55}$ erg/s and its relatively nearby location $z=0.151$ \cite{Ugarte2022}. It was so bright that it saturated most GRB detectors. However, Insight-HXMT and GECAM-C \cite{HXMT} were able to detect the exact prompt lightcurve. 

One of the unique features of GRB 221009A, on which we focus here,  was LHASSO's observations of TeV signal \cite{Huang2022,LHAASO2023} with more than 64000 photons in the energy range of $0.2-18$ TeV. 
The fast-rising and smooth, gradually declining lightcurve of the LHASSO signal \cite{LHAASO2023}, which is very different from the prompt lightcurve \cite{HXMT,Fermi2023} indicates clearly that the TeV signal was an afterglow (compare Figs. 1 of \cite{HXMT} and  1 of  \cite{Fermi2023} with Fig. 1 of \cite{LHAASO2023}). This interpretation is consistent with earlier suggestions \cite{Fan2008,Kumar2009,Kumar2010,Ghisellini2010} that most of the Fermi-LAT detected GRBs' GeV emission is an afterglow and that the TeV emission observed from GRB~190114C was from an afterglow \cite{Derishev2019,Magic2019}. 

Interestingly, the  kinetic energy of the afterglow (both as implied from the TeV emission itself \cite{DerishevPiranMNRAS2024} and from the later ($\sim 3000$ s) X-ray afterglow observations  \cite{O'Connor2023}) was significantly lower than the one of the prompt emission \cite{HXMT} (and comparable to the one observed in other TeV GRBs \cite{Derishev2019,Derishev2021}). This suggests that the extreme intrinsic properties of the prompt emission probably arose due to an extremely narrow jet \cite{HXMT,O'Connor2023}. 

The afterglow lightcurve  rises as $t^{1.8}$ from 5 to 14 s (we measure all times relative to the onset of the main prompt emission at ${T_*}=226$ s\ after the Fermi trigger\cite{LHAASO2023}).
The spectrum at this stage (see Fig. 2 of \cite{LHAASO2023}) is rather similar to those  taken 
within the intervals $14-22$ s in which the lightcurve is almost flat and $22-100$ in which the lightcurve declines as $t^{-1.115}$ (see Figs.~2 and~3b in~\cite{LHAASO2023}).  Correspondingly the lightcurves at different energies are similar (see their Figs.~4 and~S7).

\subsection{LIV limits}

Both the highest reported (7 TeV) and the lowest energy (0.2 TeV) photons arrived by 14 s, the end of the first  time interval. This puts a limit on the TOF difference between those two photons: 
\begin{equation}
\Delta t_{\rm obs}  (0.2-{7} ~{\rm TeV}) \leqslant 14 (9)\  {\rm s} \ ,
\label{lfobs}
\end{equation}
where the limit in bracket takes into account the fact that this interval begins only at 5 s. Comparing the LIV TOF relation, Eq. (\ref{eq:TF}), with the observed limit we obtain {$|\delta t_{{\rm LIV,}z=0.151}(0.2-7 ~{\rm TeV})| \leqslant  \Delta t_{\rm obs}  (0.2-{7} ~{\rm TeV})$}: 
\begin{alignat}{2}
\eqgone^{({\pm})} &\geqslant {2.8\; (4.3)}, &\qquad n&=1\ ,  \nonumber\\ 
\eqgtwo^{({\pm})} &\geqslant {5.1\; (6.4)} \times 10^{-8}, & n&=2 \ \label{eq:limits1},
\end{alignat}
where the values in  brackets reflect the stronger estimate for $\Delta t_{\rm obs}$. Strictly speaking, these limits could be affected by intrinsic spectral variations. 
As energy dependent TOF delays changes the momentarily observed spectra relative to the emitted one, it is possible in principle that spectral variability will mask LIV effects.
However, as discussed earlier, there were no significant spectral variations during this period, suggesting that intrinsic spectral variations aren't significant. Thus, Eq.~(\ref{eq:limits1}) should give correct order-of-magnitude estimate of $\eqgn^{(\pm)}$.

We can now refine these limits. In \cite{LHAASO2023} it was shown that the observed {spectral} flux, [erg$^{-1}$ cm$^{-2}$ s$^{-1}$], for $t = 5-100\,$s can be fitted by the form: 
\begin{multline}
F(E,t) = A \left( \frac{E}{\rm TeV} \right)^{-\gamma} e^{-E/E_{\rm cut}} \\
\times \left[ \left( \frac{t}{t_{\rm b}} \right)^{-\alpha_1\omega} + \left( \frac{t}{t_{\rm b}} \right)^{-\alpha_2\omega}\right]^{-1/\omega}.
\label{eq:formNoLIV}
\end{multline}
In this time range one can safely consider $\gamma$ and $E_{\rm cut}$ as constants. They, as well as $A$, $t_{\rm b}$, $\alpha_{1,2}$, and $\omega$, could be fitted to the observed lightcurve and spectra of the GRB~221009A afterglow. 

To take account of LIV TOF, we modify the {flux energy dependence by} shifting the arrival of photons according to Eq.~(\ref{eq:TF}) 
\begin{equation}
{F_{\rm LIV}}(E,t;\eqgn^{(\sigma)}) = F\bigl(E,t+\delta t_{LIV,z=0.151}(E,\eqgn^{(\sigma)})\bigr).
\label{eq:form}
\end{equation}

\begin{figure}[h]
\includegraphics[width=\columnwidth]{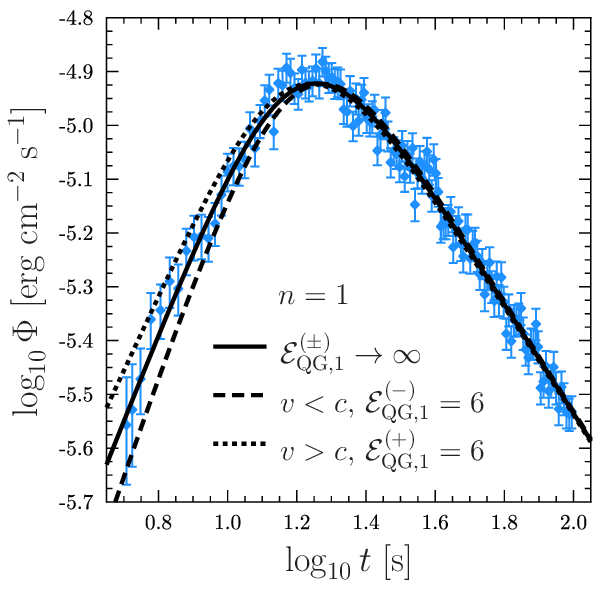}\\
\includegraphics[width=\columnwidth]{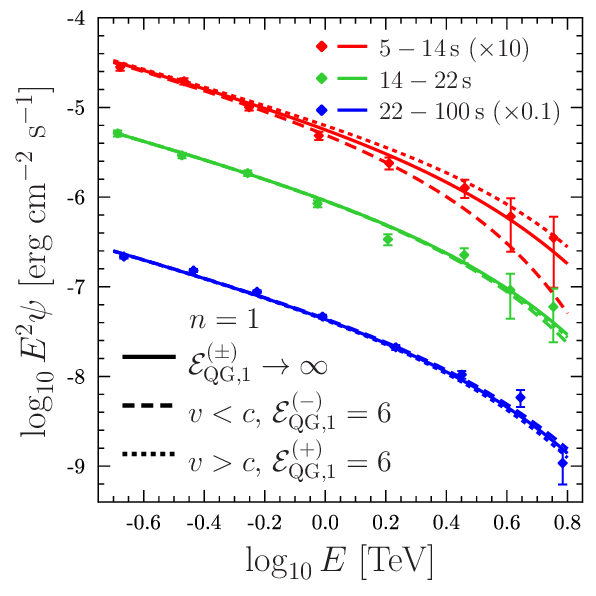}
\caption{
LHAASO data for GRB~221909A \cite{LHAASO2023} and its fitting with  $n=1$ LIV,.
\textit{Top:} The lightcurve, $\Phi(t)$,  fitted by Eq.~(\ref{eq:lcurveToFit}). \textit{Bottom:} The Spectra $\psi(E) \times E^2$ for three time bins fited by Eq.~(\ref{eq:specToFit}). The error bars reflect $1\sigma$.
The solid, dashed, and dotted lines  correspond to {$\eqgone^{(\pm)}\to\infty$, $\eqgone^{(-)}=6$, and $\eqgone^{(+)}=6$}. The latter demonstrate that $|\eqgone^{(\pm)}|\gtrsim 6 $.
}
\label{fig:1}
\end{figure}

\begin{figure}[t]
\includegraphics[width=\columnwidth]{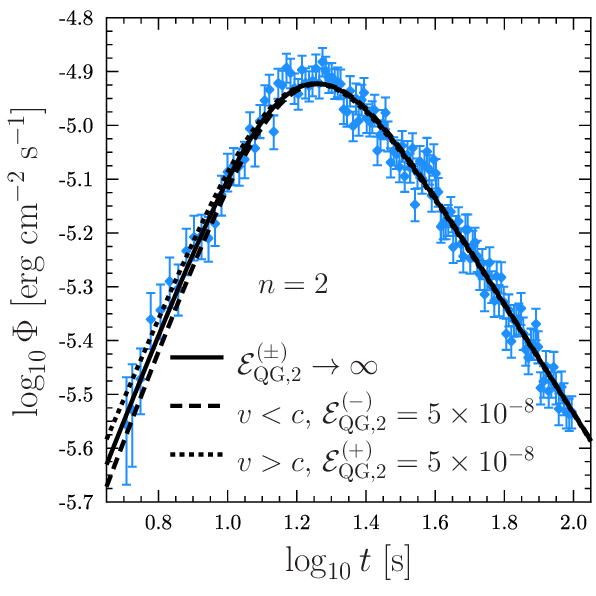}\\
\includegraphics[width=\columnwidth]{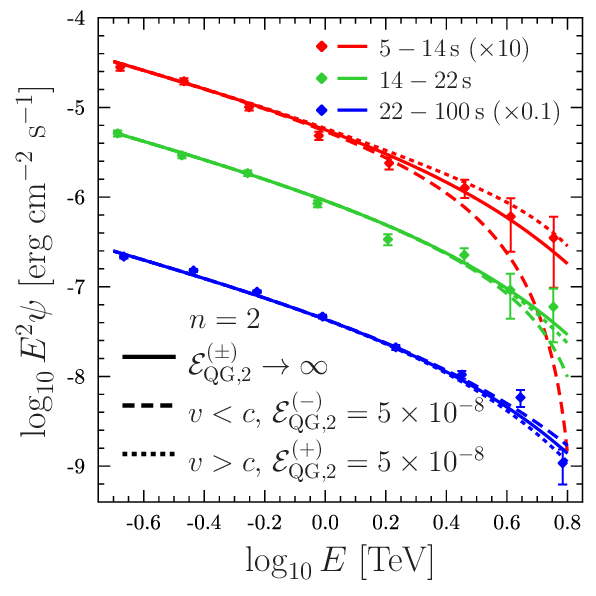}
\caption{{The same as in Fig.~\ref{fig:1} but for the quadratic LIV model, $n=2$}. The solid, dashed, and dotted lines correspond to {$\eqgone^{(\pm)}\to\infty$, $\eqgtwo^{(-)}=5\times 10^{-8}$, and $\eqgtwo^{(+)}=5\times 10^{-8}$}. }
\label{fig:2}
\end{figure}

We fit now this functional form to the LHAASO lightcurve {(Fig.~3 of \cite{LHAASO2023}; top panes in Figs.~\ref{fig:1} and~\ref{fig:2} here)} and spectrum during the first three time-bins: $5-14\,$s, $14-22\,$s, and $22-100\,$s (using data points from tables provided in \cite{LHAASO2023}), minimizing the combined $\chi^2$. We use as fitting parameters ${\sigma}/\eqgone^{{(\sigma)}}$ and ${\sigma} (10^{-8}/\eqgtwo^{{(\sigma)}})^2$ whose best-fit {values are} around zero, rather than $\eqgone$ and $\eqgtwo$ whose best-fit values, in case of no LIV, would be infinite. We have assumed that the spectrum hasn't changed during the time we consider $5-100\,$s and that the data  errors are Gaussian. 
To fit the lightcurve, $\Phi(t)$, we integrate the spectral flux~(Eq.~\ref{eq:form}) over the same energy range as in Fig.~3 of \cite{LHAASO2023}, i.e. $0.3-5\,$TeV:
\begin{equation}
\Phi_{\rm LIV}(t,{\eqgn^{{(\sigma)}}}) = \int_{0.3\,\text{TeV}}^{5\,\text{TeV}} F_{\rm LIV}(E,t;{\eqgn^{{(\sigma)}}}) E\,\mathrm{d}E.
\label{eq:lcurveToFit}
\end{equation}
To fit the spectra, $\psi^{(i)}(E)$ in each time bin $i$, we average the spectral flux over each bin:
\begin{equation}
\psi_{\rm LIV}^{(i)}(E,{\eqgn^{{(\sigma)}}}) = \frac{C_\text{a.c.}}{t_{\text{up}\,i}-t_{\text{low}\,i}} \int_{t_{\text{low}\,i}}^{t_{\text{up}\,i}} F_{\rm LIV}(E,t;{\eqgn^{{(\sigma)}}})\,\mathrm{d}t.
\label{eq:specToFit}
\end{equation}

Our method to average the spectral flux over the time bins differs from that used by \cite{LHAASO2023}. The latter uses the energy and arrival time of each photon, while the available tables provide the overall spectra and lightcurve. To take the difference into account, we introduce the phenomenological averaging-correction constant $C_\text{a.c.}$ using the same $C_\text{a.c.}$ for all three time bins. Nevertheless, as shown below, our results are consistent with those of \cite{LHAASO2023}.


Figs.~\ref{fig:1} and~\ref{fig:2} illustrate how fitting of the GRB~221009A TeV afterglow  constrains the LIV parameters. Solid lines display the best fit with no LIV ($\eqgn^{{(\pm)}}\to \infty$). Dashed and dotted lines show the results if we turn on sub- and superluminal LIV, respectively, using the same values for all other parameters as for the solid lines. The effect of LIV is two-fold. Subluminal LIV (dashed lines) suppresses the rising phase of the lightcurve and makes the spectrum during this phase ($5-14\,$s) softer. Other effects---intensifying the lightcurve in the fading phase, softening the spectrum at the maximum brightness phase ($14-22\,$s), and hardening the spectrum in the fading phase ($22-100\,$s)---are much less pronounced. For superluminal LIV, all {effects} but one are inverted (the maximum brightness spectrum is softened in both cases). In both cases, too small $\eqgn^{{(\pm)}}$ values give inconsistent discrepancies between the model curves and the data. While for $n=1$, the deviation from the lightcurve dominates, for $n=2$, the deviation from the spectral shape is the most restrictive.

\begin{table}[t]
\caption{\label{tab:table1}%
Best-fit parameters for the model 
given by Eq. \ref{eq:form} and a combined $\chi^2$ fit for both the lightcurve and the spectra of the afterglow of GRB 221009A in the time region $5-100\,$s. The {observational data is} taken from \cite{LHAASO2023}. If not specified differently, a fit parameter is dimensionless.}
\begin{ruledtabular}
\begin{tabular}{c|cc}
   & {Best fit} &   95\% confidence interval\\
\colrule
$A$\footnotemark[1]         & $8.1$   & $6.5$ --- $9.8$ \\
$C_{\rm a.c.}$              & $0.25$  & $0.24$ --- $0.26$ \\
$t_{\rm b}\footnotemark[2]$ & $16.0$     & $14.3$ --- $17.6$ \\
$\alpha_1$  &   $1.6$  &  $1.1$ ---   $2.0$  \\
$\alpha_2$  & $-1.02$  &   $-1.08$ --- $-0.95$  \\
$\omega$    &   $1.5$  &     $0.8$ ---   $2.1$  \\
$\gamma$    &   $2.93$ &    $2.84$ ---  $3.02$  \\
$E_{\rm cut}$\footnotemark[3] & $3.1$ &  $2.1$ --- $4.1$ \\
\colrule
$\eqgone^{{(\sigma)}}$      & ---\footnotemark[4]  &  
{$\eqgone^{(-)} \geqslant 5.9\ ; \eqgone^{(+)} \geqslant 6.2$} \\
$\eqgtwo^{{(\sigma)}}$ & ---\footnotemark[5] &  
{$\eqgtwo^{(-)} \geqslant 5.8 \times 10^{-8} \ ; \ \eqgtwo^{(+)} \geqslant 4.6 \times 10^{-8} $} \\
\end{tabular}
\end{ruledtabular}
\footnotetext[1]{[$10^{-6}$ erg$^{-1}$ cm$^{-2}$ s$^{-1}$]}
\footnotetext[2]{[s]}
\footnotetext[3]{[TeV]}
\footnotetext[4]{The formal fit result is $\sigma/\eqgone^{(\sigma)}= 0.005\pm 0.083$.}
\footnotetext[5]{The formal fit result is $\sigma(10^{-8}/\eqgtwo^{(\sigma)})^2 = -0.009\pm 0.019$.}
\end{table}

Results of simultaneous fitting of all nine parameters in Eqs.~(\ref{eq:formNoLIV})---(\ref{eq:specToFit}), eight for the non-violated spectral flux~(\ref{eq:formNoLIV}) and an additional one for LIV, are presented in Table~\ref{tab:table1}. They are consistent with those calculated in~\cite{LHAASO2023} when LIV is not taken into account. The  very large best-fit values for both $\eqgone^{{(\sigma)}}$ and $\eqgtwo^{{(\sigma)}}$ are consistent with infinity. For $\eqg^{{(\pm)}} \to \infty$ our results are consistent  with those of \cite{LHAASO2023}. The ranges $\eqgone^{(-)} < 5.9$ and $\eqgone^{(+)} <6.2$ and the ranges $\eqgtwo^{(-)} < 5.8 \times 10^{-8}$ and $\eqgone^{(+)} < 4.6 \times 10^{-8}$ are ruled out at 95\% confidence level. Best-fit values and confidence intervals for other parameters are almost the same for both fitting the data with linear LIV, quadratic LIV, and no LIV at all. In all three cases, the minimal $\chi^2 \approx 164$. Given the number of data points 157 and 9 independent parameters (8 for no LIV), the reduced $\chi^2$ values are $1.10-1.11$, signifying a statistically acceptable fit. The limits on $\eqgone^{(\pm)}$ we find here are comparable to the strongest limits obtained so far from GRB 090510. The limits on $\eqgtwo^{(\pm)}$ are the strongest obtained 
with the time-of-flight method
so far.
\footnote{While this paper was under review, similar results, $\eqgtwo^{(\pm)}>5.7\times 10^{-8}$, were reported by \cite{LHAASO2024LIV}.}

In the language of  Standard Model Extension (SME)  \cite{SMEreview,Kostelecky2009}  our $n=2$ limit, that corresponds to SME $d=6$ operators, yields (using the right ascention and declination of GRB~221009A~\cite{Fermi2023} and the notations of \cite{Kostelecky2009}):  
$    \sum_{jm}  {}_{0}Y_{jm}(288^\circ\!\!.\,26, 19^{\circ}\!\!.\,77)\, c_{(I)jm}^{(6)}$ is in the range $(-3,\, 4.8) \times  10^{-24}~ {\rm GeV^{-2}} $. The isotropic SME bound is $10^{-23}~ {\rm GeV^{-2}}<c_{(I)00}^{(6)} < 1.7 \times 10^{-23} ~{\rm GeV^{-2}}$. This result improves by a factor of a hundred the limits of \cite{KosteleckyMewes2008,Kostelecky2009}. Note that within the SME approach birefringence yields a better limit than TOF for $d=5$ (corresponding to $n=1$). 

\section{A comparison with LIV limits from GRBs 090510 and 190114C }
\label{sec:090510}
By now, various limits on TOF LIV have been obtained from different GRBs \cite{Ellis2006,Rodfiguez2006,Bernardini2017,Ellis2019,090510,Ghirlanda,Vasileiou2013}. We compare  our results to earlier limits obtained from two GRBs: GRB~090510 and the TeV-GRB~190114C. One of the remarkable implications of LHASSO observations of GRB 221009A was that the TeV emission is an afterglow. This simplifies the interpretation of the data as the afterglow has a much simpler, albeit typically broader, lightcurve, as compared with the prompt emission. Interpretation of the spectra (although we don't use it here) is also easier. Additionally, both GRBs displayed very high energy emission (30 GeV for GRB~090510 and a few TeV for GRB~190114C) and limits obtained just from the low-energy gamma-rays cannot exceed a few  hundredths of the Planck scale \cite{Rodfiguez2006a}. Therefore, we focus, on the comparison with the interpretation of GRB 090510 data that was based on the assumption of an afterglow source \cite{Ghirlanda} and on the limits obtained from GRB 190114C, which were also based on an afterglow model. These are also the best limits available so far. 

GRB 090510 was a bright short burst with one of the first Fermi-LAT detection of GeV photons \cite{Fermi2010}. In \cite{090510}, the Fermi team obtained several limits, using different assumptions on the origin of the GeV photons and their association with the lower energy ones. Shortly afterward, it was proposed that the GeV emission arose from an afterglow \cite{{Kumar2009,Kumar2010,Ghisellini2010}}. A crude TOF LIV limit  calculated based on this assumption gave \cite{Ghirlanda}: 
$\eqgone^{(\pm)}>4.7 (6.7) $, for a rise time estimate of  $\Delta t_{\rm obs} =0.217$ s ($\Delta t_{\rm obs} = 0.15$ s), ignoring  the question of superluminal vs. subluminal LIV correction.
While in \cite{Vasileiou2013} the authors didn't consider specifically an afterglow model, their results, which focus on what they call the ``main" peak, indirectly make such an assumption. Hence, we consider these values here. Their 95\% confidence range (using their Maximal Likelihood method---other methods give comparable limits) are 
$\eqgone^{(-)} > 11$ and $\eqgone^{(+)} > 5.2$, and $\eqgtwo^{(-)} > 0.7 \times 10^{-8}$ and $\eqgtwo^{(+)} > 0.77 \times 10^{-8}$. 

It is interesting to compare the two data sets (see Table \ref{tab:table2}). In GRB 090510 the observed time scale is  shorter by a factor of $\sim 100$, and the distance is larger by a factor of $\sim 6$. On the other hand, in GRB 221009A, the highest energy photon is more energetic by a factor of $\sim 200$. This suggests that 090510 should give comparable but slightly better limits for $n=1$. The larger number of photons somewhat improves the balance towards 221009A. As the energy scale is much more important for $n=2$, in this case, the 221009A limits are much more significant.  

\begin{table}
\caption{\label{tab:table2}%
A comparison of with LIV TOF limits from GRBs 090510, 190114C, and 221009A.}
\begin{ruledtabular}
\begin{tabular}{cccc}
\textrm{GRB}&
\textrm{090510\footnote{From \cite{Vasileiou2013} with ML method. }}&
{\textrm{190114C}}&
\textrm{221009A}\\
\colrule
\textrm{Red Shift} & 0.903 & 0.425 &  0.151\\
$\Delta E ~~{\rm [TeV]}$ & $10^{-4}$ --- 0.03& 0.3 --- 1 & 0.2 --- {7} \\
$\Delta T_{\rm obs}~{\rm [s]}$&0.15 --- {0.217} & 30 --- 60 & 9 --- {14}  \\
\colrule
$\eqgone^{{(\sigma)}} $&{11$^-$ 5.2$^+$ }   & {0.23$^-$ 0.45$^+$ } & {5.9$^-$ 6.2$^+$ } \\
$\eqgtwo^{{(\sigma)}}/10^{-8}$ &{0.7$^-$ 0.77$^+$}   & {0.46$^-$ 0.52$^+$}  & {5.8$^-$ 4.6$^+$  }\\
\end{tabular}
\end{ruledtabular}
\end{table}

GRB 190114C was the first event with reported TeV emission \cite{Magic2019}. The MAGIC telescope began observations of this event 62 s after the trigger, detecting photons above 0.2 TeV up to  $\sim 1100$ s. Unfortunately, the MAGIC observations caught the afterglow already in the declining phase. Thus, they provide only an upper limit on the afterglow peak, leading to a rather large $\Delta t_{\rm obs}$ value. Analysis of the lightcurve and spectrum of this event \cite{Acciari} gives lower limits of $\eqgone^{(-)}> 0.23 (0.475)$ and $\eqgone^{(+)} > 0.45$ (We present the results in units of $\mpl$ while \cite{Acciari}  present them in units of $10^{19}$ GeV) where the value in brackets reflects a less conservative (model dependent) assumption about the intrinsic lightcurve. These limits are not competitive with those from 090510 or from 221009A. While the photons' energy was comparable to those in 221009A and the redshift was larger, the lack of precise estimate of $\Delta t_{\rm obs}$ reduced the effective limit.  Because of the higher energy of the observed photons,  the $n=2$ limits are comparable: $\eqgtwo^{(+)} > 0.46 \times 10^{-8}$ and $\eqgtwo^{(-)}>0.52 ~(0.6) \times 10^{-8}$ ({We present the results in units of $\mpl$ while \cite{Acciari}  present them in units of $10^{10}$ GeV }). These limits are also comparable to those of \cite{Mk501} that are based on the strong TeV flare of Mrk 501. Our limits from 221009A are almost an order of magnitude better.

\section{Conclusions}
\label{sec:conclusions}

We have found new limits on LIV, as measured by time of flight, from the TeV  lightcurve and spectra reported by  LHAASO for GRB 221009A. For $n=1$ we find that the minimal energy for LIV is a few Planck masses for both superluminal and subluminal modes. For $n=2$ our limits are a few times $10^{-8} \mpl$. It is not surprising not to find significant LIV modifications at this energy scale.  

Following this work, similar results were reported by the LHAASO collaboration \cite{LHAASO2024LIV}. Using the full raw data,  they obtained, as expected, slightly more restrictive constraints, $\eqgone^{(-)}>8.2$, $\eqgone^{(+)}>9.0$, and $\eqgtwo^{(\pm)}>5.7\times 10^{-8}$ at 95\% level.

As GRB 221009A is much nearer than GRB 090510 and its variability time is much longer, our $n=1$ limits are comparable to those obtained by \cite{090510,Ghirlanda,Vasileiou2013} from GRB 090510, even though
the observed energy for the GRB 221009A photons is much higher. The higher energy plays a more important role for $n=2$, and in this case, our limits are almost an order of magnitude higher than previous ones.  Still, the $n=2$ limits are far from the Planck scale. Unfortunately, reaching a more significant limit on $n=2$ is not within reach using photons, as the EBL  absorbs higher energy photons. We may have to wait for a combined electromagnetic and very high energy neutrino flare to explore that \cite{jacob2007}.

An intrinsic problem of the time-of-flight method, when based on a single source, is that it is impossible to distinguish between intrinsic and LIV-induced spectral variation. 
A way to distinguish between the two is by a joint analysis of different GRBs from different redshifts \cite{Ellis2019}. The LIV time delay has a clear redshift dependence that will stand out.  Such an analysis \cite{Ellis2019} gave a robust lower limit of $\eqgone^{{(\pm)}} > {\rm a ~few} \times 0.01$. These authors warn against using bounds obtained from a single GRB as those are prone to possible intrinsic spectral variation. However, we have introduced here a data point at a different redshift. The combination of the independent limits from different redshits is much more powerful than the individual ones. It ensures that for $n=1$ the typical scale for LIV-induced photon speed modification happens at least on a scale of a few Planck masses.

\acknowledgements
TP thanks Evgeny Derishev for fruitful discussions on GRB 221009A, the members of the COST CA18108 collaboration for illuminating discussions on LIV and an anonymous referee for helpful remarks. 
This research was supported by an Advanced ERC grant MultiJets and by ISF grant 2126/22.


\providecommand{\noopsort}[1]{}\providecommand{\singleletter}[1]{#1}%

\end{document}